\documentclass{article}

\usepackage{pslatex}

\usepackage{amssymb}
\usepackage[dvips]{epsfig}
\usepackage{graphicx}

\title{Acoustic characterization of Hofstadter butterfly with resonant scatterers}

\author{O. Richoux\thanks{Laboratoire d'Acoustique de l'Universit\'e du Maine UMR CNRS 6613 - Avenue O. Messiaen, 72085 Le Mans Cedex 9, France. E-mail : olivier.richoux@univ-lemans.fr} \and V. Pagneux \thanks{Laboratoire d'Acoustique de l'Universit\'e du Maine UMR CNRS 6613 - Avenue O. Messiaen, 72085 Le Mans Cedex 9, France. E-mail : vincent.pagneux@univ-lemans.fr}}
\date{1 July 2002}

\begin{document}

\maketitle

\begin{abstract}

We are interested in the experimental characterization of the Hofstadter butterfly by means of acoustical waves. The transmission of an acoustic pulse through an array of $60$ variable and resonant scatterers periodically distribued along a waveguide is studied. An arbitrary scattering arrangement is realized by using the variable length of each resonator cavity. For a periodic modulation, the structures of forbidden bands of the transmission reproduce the Hofstadter butterfly. We compare experimental, analytical, and computational realizations of the Hofstadter butterfly and we show the influence of the resonances of the scatterers on the structure of the butterfly.

\end{abstract}

\section{Introduction}

The Hofstadter butterfly is well known since the work of Hofstadter on the electron transmission through a two-dimensional ordered lattice perturbed by a perpendicularly applied uniform magnetic field \cite{Hofstadter}. Few experimental attempts have been made to observe signatures of this phenomenon in two-dimensional electron systems \cite{Schlosser96, Ensslin97,Gerhardts91}. Recently, an experimental electromagnetic realization of the Hofstadter butterfly was done by studying the transmission of microwaves through an array of $100$ scatterers inserted into a waveguide when the modulation length of this unidimensional periodic lattice changed \cite{Kulh}. These authors used cylindrical scatterers introduced into a rectangular microwave waveguide. In this study, the stopbands were produced by Braggs reflection on the transmission spectra and the illustration of the Hofstadter butterfly exhibited self-similar structures. 

In the present paper, we study the transmission of an acoustic pulse through a lattice composed of an assembly of variable Helmholtz resonators that are the scatterers \cite{Pierce}. A particularity of this work is the use of scatterers that can exhibit resonant scattering at wavelength much larger then the scatterers (well known examples of Helmholtz resonators are the bottle of wine that have resonances for frequencies corresponding to wavelengths of the order of the meter). The resonators are uniformly distributed along a cylindrical waveguide. Each Helmholtz resonator is constituted by two cylindrical tubes which play the role of neck and cavity of the resonator. The length of the Helmholtz cavity is a variable parameter and the lattice spacing constant $d$ between each resonator is fixed.

The propagation of sound wave into a tunnel with a periodic array of Helmholtz resonators has been already studied \cite{Sugimoto} and different kinds of stopband appear in the transmission coefficient. In addition to Bragg stopbands caused by the spatial periodicity of the lattice, resonances of the Helmholtz resonator create other "stopping bands" called scatterer stopbands which are non-existent in \cite{Kulh}. We show that a realization of the Hofstadter butterfly is obtained experimentally with acoustic waves. The effects of the scatterer stopbands are examined. Experimental work is completed by an analytical and a numerical analysis, and all show self-similar structure.

\section{Theory of the propagation into a lattice}

\begin{figure}[h!]
\centering
\includegraphics[width=12cm]{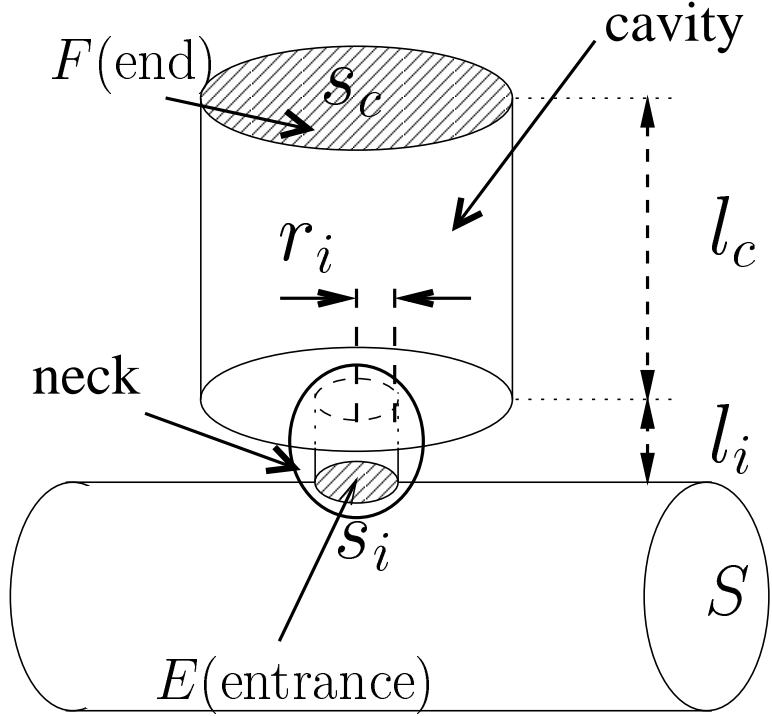}
\caption{\label{fig2} Description of the Helmholtz resonator.}
\end{figure}

We study a monochromatic pressure wave $\tilde{p}(x,t)$ with the form $\tilde{p}(x,t)=p(x) e^{j\omega t}$ where $\omega$ is the wave pulsation. A cylindrical waveguide with Helmholtz resonators connected axially constitutes the one-dimensional lattice. In this lattice, the pressure $p(x)$ is solution of the following equation \cite{Levine},
\begin{equation}
\label{eqprop}
\frac{d^2 p(x)}{dx^2}+k^2 p(x)= \sum_{n} \delta(x-x_n) \sigma_n p(x)
\end{equation}
where $k=\omega /c$, $\sigma_n=-j \omega \rho \frac{s_i}{S} Y_n$ with $s_i$ and $S$ the derivation and the main tube areas respectively (fig. \ref{fig2}), $\rho$ is the density of air and $c$ is the celerity of the wave. $Y_n$ is the  admittance of the n$^{th}$ resonator placed at $x=x_n$, defined by 
$
Y_n=v(x_n)/p(x_n)=v_n/p_n
$
where $p(x_n)$ and $v(x_n)$ are  respectively the pressure and the acoustic velocity at $x=x_n$. The solutions of eq. (\ref{eqprop}) are found by the transfer matrix method. In the region $x_n \le x \le x_{n+1}$, which determines the n$^{th}$ cell, the solutions of eq. (\ref{eqprop}) are separated into two plane waves with opposite propagation direction associated with the amplitudes $A_n$ and $B_n$: 
$
p(x)=A_n e^{jk(x-x_n)}+ B_n e^{-jk(x-x_n)}.
$

The propagation through one cell is described by
\begin{equation}
 \left( \begin{array}{c} A_{n+1} \\ B_{n+1} \end{array} \right)=
M_n \left( \begin{array}{c} A_n \\ B_n \end{array} \right) =
\left( \begin{array}{cc} (1-\frac{\sigma_n}{2jk})e^{-jkd_n} & \frac{\sigma_n}{2jk}e^{jkd_n} \\
-\frac{\sigma_n}{2jk}e^{jkd_n} & (1+\frac{\sigma_n}{2jk})e^{-jkd_n} \end{array} \right) 
\left( \begin{array}{c} A_n \\ B_n \end{array} \right)
\end{equation}
with $d_n=x_{n+1}-x_n$. For a finite lattice which is composed of $N$ resonators, the wave propagation is described by $(A_n \, B_n)^t=\prod_{n=1}^{N} M_n (A_0 \, B_0)^t$, 
where $(A_0 \, B_0)^t$ are the amplitudes of the inlet of the lattice.

The resolution of the eigenvalue problem for the matrix $M_n$ allows us to find the structure of the transmission (allowed and forbidden bands). By setting $\lambda^{\pm}=e^{\pm jqd}$ for the eigenvalues, the dispersion relation takes the form 
\begin{equation}
\label{eq2}
  \cos (qd) = \cos(kd)+ \frac{\sigma_n}{2k}\sin(kd),
\end{equation}
where $q$ is called the Bloch wave number. The transmission of the lattice is determined by the value of $\cos (qd)$ : if $\cos (qd)$ belongs to $[-1,1]$, the wave propagates into the system; in contrast, when $\cos (qd)$ belongs to $]-\infty,-1]$ or $[1,+\infty[$, the wave is exponentially damped and the lattice is completely opaque to the wave.

\section{Determination of Helmholtz resonator admittance $Y$}

To determine the impedance of a Helmholtz resonator, we need to a relation pressure $p$ and velocity $v$ between the entrance (E) and the end (F) of the resonator (fig. \ref{fig2}). Applying classical lossless plane wave technique it is found that 
\begin{eqnarray}
\label{eq1}
\left(
\begin{array}{c}
p_E \\ v_E
\end{array} \right) 
& = & 
 \left( \begin{array}{cc}
  \cos(kl_i) & jZ_c^i \sin(kl_i) \\
  \frac{j}{Z_c^i} \sin(kl_i) & \cos(kl_i) 
  \end{array} \right) 
   \left( \begin{array}{cc}
  1 & jZ_c^i k \Delta l \\
  0 & 1 
  \end{array} \right) 
\nonumber \\
& \times &
  \left( \begin{array}{cc}
  \cos(kl_c) & jZ_c^c \sin(kl_c) \\
  \frac{j}{Z_c^c} \sin(kl_c) & \cos(kl_c) 
  \end{array} \right) 
  \left( \begin{array}{c}
  p_F \\ v_F
  \end{array} \right)
\end{eqnarray}
where $Z_c^i$ and $Z_c^c$ are respectively the characteristic impedance of neck and cavity, $l_i$ and $l_c$ are respectively the length of the neck and the length of the volume cavity, and subscripts $E$ and $F$ refer to entrance and end of the resonator. $\Delta l$ represents the added length to the neck due to the discontinuity of the sections. This length takes generally the form \cite{Dubos} 
$
\Delta l = \Delta l_i + \Delta l_c = r_i(\frac{8 \pi}{3}+  \frac{8 }{3 \pi} + o(\frac{s_i}{S}))
$
where $r_i$ is the radius of the neck.

Because the Helmholtz resonator is closed, the boundary condition at the end of the resonator is $v_F=0$. Eliminating $p_F$ in the eq. (\ref{eq1}), the acoustic impedance $Z=p_E/v_E$ seen from the entry (at the point E) is given by the following relation :
\begin{equation}
\label{impedance}
Z=\frac{1}{Y}=\frac{\cos(kl_i) \cos(kl_c) - \frac{Z_c^i}{Z_c^c}k(\Delta l)\cos(kl_i) \sin(kl_c) -  \frac{Z_c^i}{Z_c^c}\sin(kl_i)\sin(kl_c)}{ \frac{1}{Z_c^i}\cos(kl_c)\sin(kl_i) - \frac{1}{Z_c^c} \sin(kl_i)\sin(kl_c) +  \frac{1}{Z_c^c} \cos(kl_i)sin(kl_c)}.
\end{equation}

Because of resonances, it is very important to take into account the losses. The theory with losses \cite{Zwikker} is used in the eq. (\ref{impedance}) with $k=\frac{\omega}{c_0}(1+\frac{\beta}{s}(1+(\gamma -1)\chi)$ and $Z_c = \rho c( 1+\frac{\beta}{s}(1-(\gamma -1)\chi)$ 
by setting $s=r/\delta$ where $\delta$ is the viscous boundary layer, $\chi=\sqrt{P_r}$ with $P_r$ the Prandtl number, $\beta=(1-j)/\sqrt{2}$ and where $r$ is the radius of the tube considered.

\section{Description of the experimental apparatus}

In the experiment, a cylindrical waveguide with an inner radius $r=2.5 \, 10^{-2}$ m was used. $60$ Helmholtz resonators were periodically located along this waveguide and the lattice spacing between each resonator was $d=0.1$ m. Each resonator was composed by a neck (cylindrical tube with an inner area $S_i=7,85 \, .10^{-5} \, m^2$ and a length $l_i=2 \, .10^{-2}$ m) and a variable length cavity (cylindrical tube with an inner area $S_c=7,85 \, .10^{-3}\, m^2 $ and a maximum length $l_c=16.5 \, .10^{-2}$ m) as described in the fig. \ref{fig2}. The sound source was a loudspeaker with special design placed at one end of the main tube . Two microphones (BK 4136) measured the pressure at each end of the cylindrical waveguide. We measured in the frequency range where only the first mode can propagate, starting from $0$ Hz up to the first cutoff frequency $f_{c01}=4,061$ kHz, where the propagation of the first higher mode becomes possible.

For the experimental realization of the butterfly a periodic modulation of the lengths of the cavity resonators was applied with the period length as a parameter. The modulation of this lattice is given by the variation of the cavity length of the n$^{th}$ resonator 
\begin{equation}
  l_n=l_c\cos(2 \pi n \alpha - \alpha_0)
\end{equation}
where $\alpha=\frac{p}{m}$, with $p,\, m \in \mathbb{N}$. We replace this cosinusoidal variation by a rectangular one (in the same manner as \cite{Kulh}):
\begin{equation}
  l_n = \left\lbrace \begin{array}{ccc}
        0 & \mbox{ for } & \cos(2 \pi n \alpha - \alpha_0) \leq 0 \\
        l_c & \mbox{ for } & \cos(2 \pi n \alpha - \alpha_0) > 0.
        \end{array} \right.
\end{equation}
In the experiment we choose $l_c=16.5 \,10^{-2} \, m$ and $\alpha_0=0$. It is important for the experiment to use a rectangular setup and not a sinusoidal otherwise the stopbands would destroy the butterfly.

\section{Signal analysis}

To investigate the experimental results, we used a time-frequency method. It maps the time domain signal (the square of the amplitude) in the time-frequency plane. In this way, it is possible to know the arrival time of each frequency present in the input signal and to determine a "transfer function" of the acoustical system. The Wigner-Ville \cite{Flandrin} distribution defined as 
$$
W_z(t,f)=\int_{-\infty}^{+\infty} z_x(t+\frac{\tau}{2})z_x^*(t-\frac{\tau}{2})e^{j2\pi f \tau} d\tau
$$
where $z_x(t)$ is the analytical signal \cite{Gabor} (associated to the real signal $z(t)$) is certainly the most widely studied time-frequency methods. To reduce the cross terms (interferences) between different components of the signal, we used the Pseudo Wigner-Ville distribution PW given by :
$$
PW_z(t,f)=\int_{-\infty}^{+\infty} |h(\tau)|^2 z_x(t+\frac{\tau}{2})z_x^*(t-\frac{\tau}{2})e^{j2\pi f \tau} d\tau
$$
where $h(\tau)$ is the temporal rectangular window with a length $\tau$. This method gives the acoustical response of the lattice in the time-frequency plane by using a source pulse. We determined the transmission spectra by detecting the maximum of energy for each frequency upstream and downstream the lattice. The transmission was deduced from the "transfer function" (in energy) defined as the fraction of transmitted and incident energy.

\begin{figure}[h!]
\centering
\includegraphics[width=12cm]{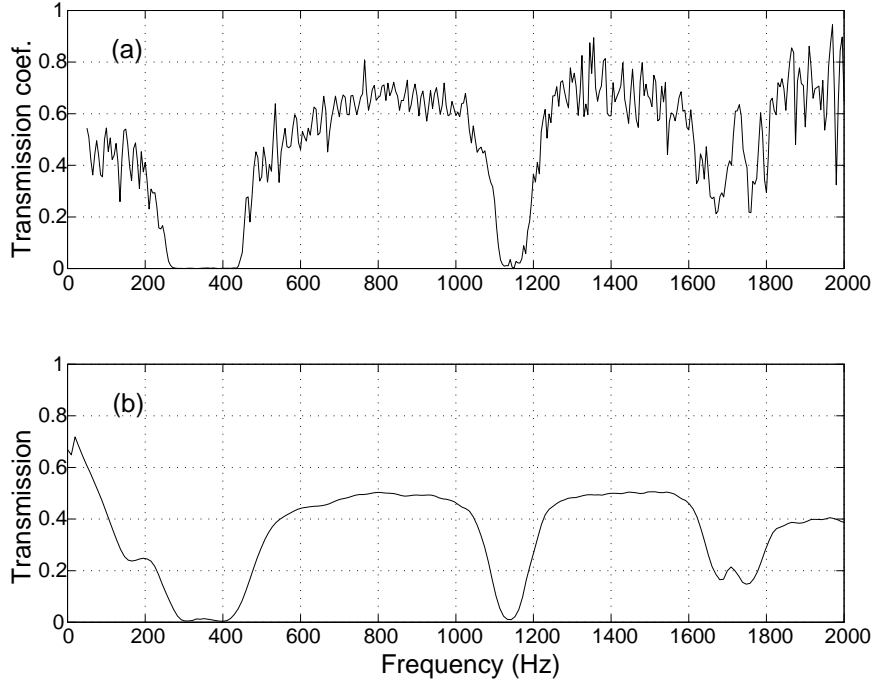}
\caption{\label{fig3} Comparison between two different signals processing for $\alpha=0.0125$. (a) Use of a Fourier transform and a shirp source . (b) Use of Wigner-Ville distribution and a pulse source.}
\end{figure}

Using a pulse source, this method avoids seeing, in the transmission spectra, oscillations phenomena caused by the finite number of cells \cite{Griffiths92}. The pulse source prevents this phenomena, which would destroy the Hofstadter butterfly structure, when the propagation time along a cell is greater than the pulse duration \cite{Richoux}. The fig. \ref{fig3} shows a comparison between these two sorts of processing method : a classical way (Fourier transform) using a shirp source and the previous one using the pseudo distribution of Wigner-Ville and a pulse source. It illustrates the oscillations in the transmission coefficient and shows the influence of the source on the transmission spectra.

\section{Analytic determination of the Hofstadter butterfly}

The periodic rectangular modulation implies only two values for the cavity length of each Helmholtz resonator. When this length vanishes, the resonator effects on the propagation are completely invisible in the frequency range of interest. When $\alpha$ is written as $\alpha=1/m$, the lattice is periodic with a period length $d'=d/\alpha=md$. In this case, analytical points of the Hofstadter butterfly can be found in the $(f,\alpha)$ plane.

This new period $d'$ is used in the dispersion relation (eq. (\ref{eq2})) which becomes $\cos(kd')+\frac{\sigma}{2k}\sin(kd')=\pm1$.

\begin{figure}[h!]
\centering
\includegraphics[width=12cm]{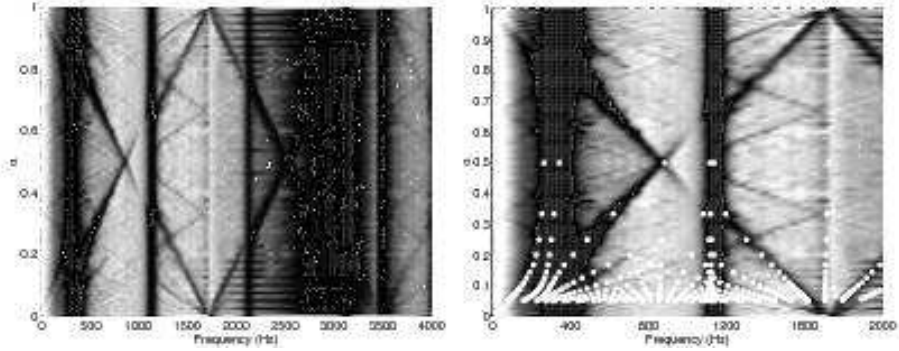}
\caption{\label{fig5} (a) Experimental transmission spectra for a periodic arrangement of scatterers with $\alpha$ ranging from $0$ to $1$ with $50$ values between $0$ and $0.5$. The upper part is obtained by reflection. (b) Zoom of the experimental transmission spectra for a periodic arrangement of scatterers with $\alpha$ ranging from $0$ to $1$ between $[0:2000]$ Hz. The white points show the analytic results from the eq. (\ref{ana1}).}
\end{figure}

\begin{figure}[h!]
\centering
\includegraphics[width=12cm]{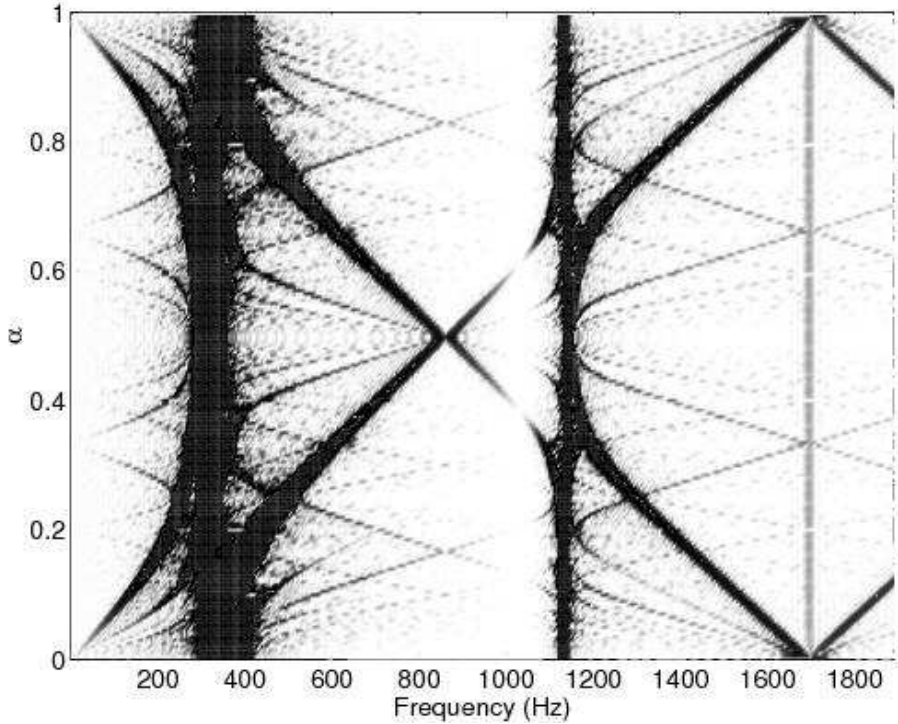}
\caption{\label{fig6} Numerical result for the transmission spectra for a periodic arrangement of scatterers with $\alpha$ ranging from $0$ to $1$ with $100$ values between $0$ and $0.5$. The upper part is obtained by reflection.}
\end{figure}

The solutions of this equation gives some points of the line which is the boundary between the passbands and the stopbands (the stopband are determined when the dispersion relation (\ref{eq2}) is greater than $1$ or lower than $-1$). By setting 
$
\tan(\theta)=-\frac{\sigma}{2k},
$
the dispersion relation is written as
$
\cos(k\frac{d}{\alpha}+\theta)=\pm \cos(\theta)
$ 
and the solutions are
\begin{equation}
\label{ana1}
k\frac{d}{\alpha}=n\pi-2\theta \mbox{ and } k\frac{d}{\alpha}=n\pi \mbox{ with } n \in \mathbb{N}. 
\end{equation}
The effects of the scatterers are present in the term $\theta$ and the resonance frequencies imply its divergence. These divergences perturb branches structure of the butterfly around these resonance frequencies and "break" the butterfly structure as presented in the fig. \ref{fig5} and \ref{fig6} for $f = 283$ Hz, $f=1120$ Hz, $f=2240$ Hz and  $f=3360$ Hz (each band corresponds to a resonance of the resonator). The resolution of the eq. (\ref{ana1}) gives different points on the $(f,\alpha)$ plane for $\alpha$'s values corresponding to $p=1$ and $m=2$ up to $20$ which are reported on the fig. \ref{fig5}b.

\section{Results}

The transfer function between the end and the beginning of the lattice, in the frequency range $[0,4000]$ Hz which is corresponding to two $\pi$-Bragg bands, was measured. We used $50$ different values for the parameter $\alpha$ between $0$ and $0.5$. 

\begin{figure}[h!]
\centering
\includegraphics[width=12cm]{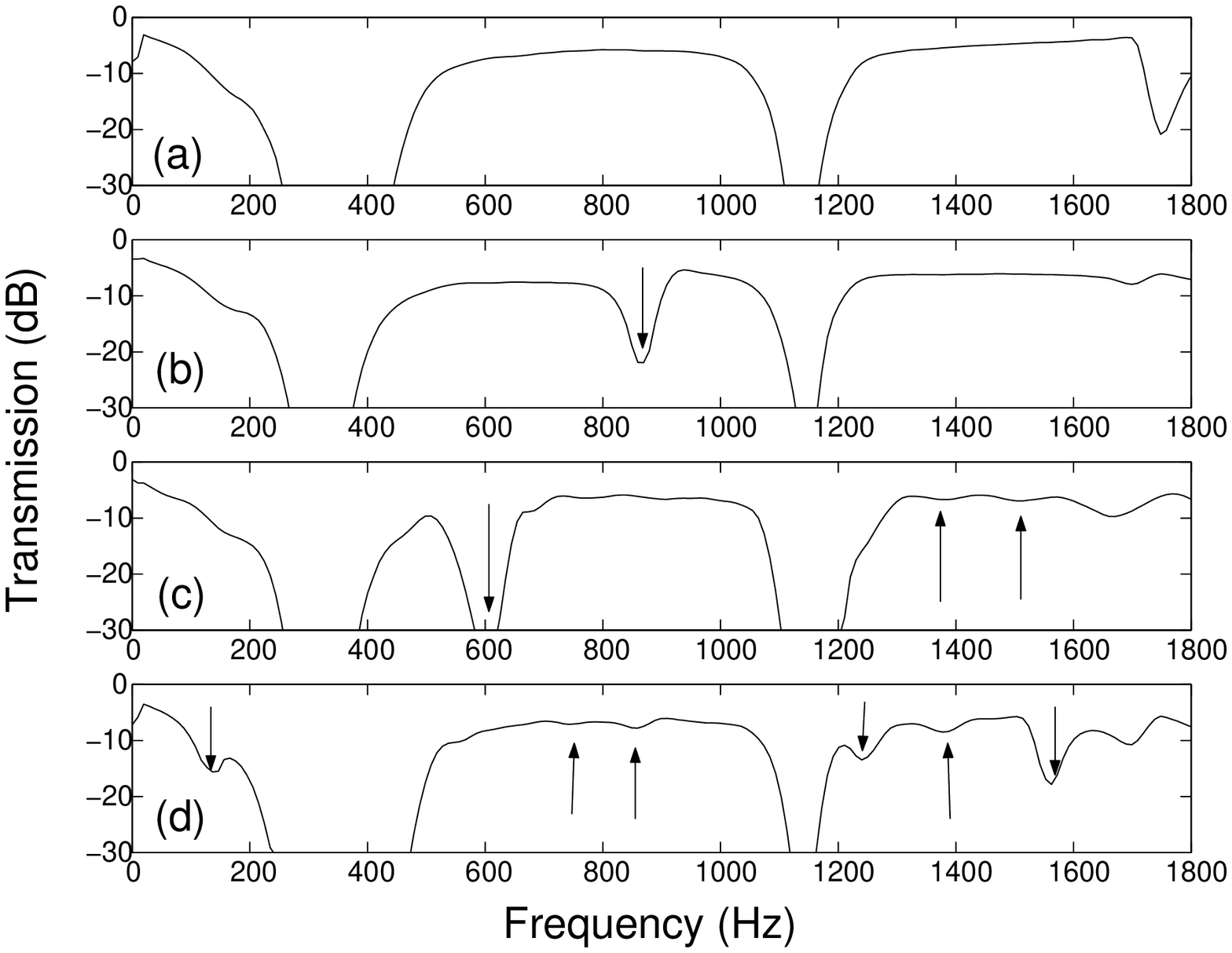}
\caption{\label{fig4} Transmission function for four different values of modulation lengths. (a) $\alpha=0$. (b) $\alpha=0.5$.  (c) $\alpha=0.325$. (d) $\alpha=0.1$. Some subbands due to the modulations are marked by arrows.}
\end{figure}

The fig. \ref{fig4} presents the transmission in the periodic case (all the resonators are identical) and for three different values of modulation length. The periodic case (fig. \ref{fig4}a) is obtained for $\alpha=0$ and shows two sorts of forbidden bands in the spectra : the first and second are due to the scatterers (Helmholtz resonators) for $283$ Hz and $1120$ Hz and the third is due to the spatial periodicity ($\pi$-Bragg stopband) at $1700$ Hz. 

The periodic modulation of the lattice involves the arrival of forbidden subbands in the transmission spectra (fig. \ref{fig4}b, c and d) which split the transmission band (allowed band) into subbands (marked with a arrow in the figure). The number of these subbands increases with $m$. The new period of the lattice is $d'=md$ so the frequency location of the subbands are now $f_b=c/(2md)$ :$1$ subband appears for $p=1$ and $m=2$ (fig. \ref{fig4}b) at $850$ Hz and theoretically $10$ subbands for $p=1$ and $m=10$ (fig. \ref{fig4}d) every $170$ Hz (this argument is not satisfactory for $p \not= 1$ as in the fig. \ref{fig4}c where $p=13$ and $m=40$). The attenuation phenomenon added to the finite dimension of the lattice and the presence of stopbands due to resonance's scatterers induce that some subbands are missing in the spectra presented in the fig.~\ref{fig4}.

To obtain a view of the Hofstadter butterfly, the transfer functions for different values of the modulation parameter $\alpha$ are plotted together on a plane building with the frequency (abscise axe) and $\alpha$ belonging to $[0,1]$ (ordinate axe). The spectra are converted to a grey scale (fig. \ref{fig5} and \ref{fig6}) and the part between $[0.5,1]$ is obtained by reflection.

In fig. \ref{fig5}a two $\pi$-Bragg bands are seen and two Hofstadter butterflies are complete. The structure of the butterflies is broken by the resonances of the resonators and some "oscillations" due to the signal processing are observed. The fig. \ref{fig5}b presents a zoom of the fig. \ref{fig5}a for a $[0:2000]$ Hz frequency range. In this figure, the structure of the Hofstadter butterfly are identified and can be compared to the simulation (fig. \ref{fig6}) obtained with $100$ values of $\alpha$. Self similar structures are observed and a lot of subbands are perceptible. The presence of scatterers stopbands seems to stop the effects of the lattice modulation in the transmission : for example, the main subband is just shifted during the both stopbands caused by Helmholtz resonators and the butterfly structure is affected by this phenomenon. The white points on the fig. \ref{fig5}b present the analytical results calculated with the eq. (\ref{ana1}). They are in agreement with the experimental results and they predict sufficiently the effect of the scatterer resonances.

The self-similar structure of the butterfly is also observed. The simulated butterfly (fig. \ref{fig6}), building with $100$ different arrangements of the scatterers, exhibits more self-similar structures than the experimental results. The depth of the self-similar structures is depending on the number of modulation used to map the Hofstadter butterfly. The simulation allows to observe with a high definition some fine details of the Hofstadter butterfly and to analyze more precisely the effects of the scatterer resonances. It seems that the presence of scatterer stopbands (the first and the second stopbands) breaks the modulation influence by diverting the different branches of the butterfly. It implies that the structure is self-similar only in the region between the scatterer stopbands. This phenomena is also observed with the experimental results for the main branch of the butterfly (fig. \ref{fig5}b). 

\section{Acknowledgments}

We are grateful to the referees for helpful suggestions.


\end{document}